# Using Alloy to model-check visual design notations

Anthony J. H. Simons, Carlos Alberto Fernández y Fernández
*Department of Computer Science, University of Sheffield*
*{A.Simons, C.Fernandez} @dcs.shef.ac.uk*

## Abstract

*This paper explores the process of validation for the abstract syntax of a graphical notation. We define a unified specification for five of the UML diagrams used by the Discovery Method and, in this document, we illustrate how diagrams can be represented in Alloy and checked against our specification in order to know if these are valid under the Discovery notation.*

## 1. Introduction

The Unified Modeling Language (UML) [1] is an eclectic set of notations for modelling object-oriented designs. Under the supervision of the Object Management Group (OMG), the notation set has grown larger, to accommodate the concerns of different stakeholders in business and industry. This has led to some criticisms regarding the open ended semantics and the lack of direction given in modelling [2].

Various attempts to formalise parts of UML include the work of the Precise UML group (pUML) [3], which aims to clarify the semantics of UML and create tools to support the rigorous analysis of UML models. Jointly with IBM, pUML submitted a Meta-Modelling Framework (MMF) [4] to the OMG as an option to the original UML metamodel. Out of this work came the desire to create an Unambiguous UML, an idea partly inspired by the Catalysis method [5]. The Unambiguous UML (2U) Consortium [6], which grew out of pUML, submitted a full proposal for UML2.0 based on a set of architectural principles.

Related work on the development of model checkers and tools for UML has been quite slow. Some examples include the Hugo tool, which compiles UML state machines into a format processed by the PROMELA model checker [7]. The USE tool (UML-based Specification Environment) [8] allows UML diagrams to be annotated with constraints written in OCL (the Object Constraint Language [9]), after which the validity of models may be checked, using predicates also written in OCL. The tool verifies model instances against explicit predicates and also implicitly against the invariants defined in the model.

A prerequisite to developing high-quality model checkers is the ability to encode model diagrams in a suitable abstract syntax, and from this to develop an abstract semantics [10]. This paper reports on a series of initial experiments conducted into converting a small object-oriented design notation into an abstract syntax, which was then submitted to the model checker Alloy [11, 12], to verify the correctness of models against the abstract syntax. The design notation is a subset of UML used by the Discovery Method [13, 14]. We demonstrate that it is possible to validate model instances of different types against an abstract syntax specification. We also show how a combination of consistent model instances yields a single, consistent abstract syntax; and the converse, that inconsistent models are rejected against the abstract syntax specification. The Alloy analyzer proves to be a tricky tool to use well in this context, and we report also on various different approaches to encoding diagram information and the tactics adopted to focus the work of the model checker.

The rest of this paper is organized as follows: section 2 discusses the relationship between UML and the notation of the Discovery Method; section 3 gives a brief explanation of the Alloy analyzer; section 4 describes the research methodology; section 5 introduces the abstract syntax model for encoding Discovery Method notations in Alloy; section 6 gives an example of usage and model-checking; section 7 gives an initial evaluation of Alloy and its usefulness for modelling diagram syntax constraints; and finally section 8 presents our conclusions. A graph depicting the Alloy abstract syntax metamodel for Discovery Method notations is also included in an appendix.

## 2. UML and the Discovery Method

Supporters of UML argue that designers benefit from being able to choose whatever diagram elements they need, with some latitude to interpret the diagrams as they see fit. Others argue that this freedom is undesirable, resulting from the lack of any unified semantics in UML [15]. In fact, Booch [16] has stated that the current UML specification does not restrict graphical formats and *"there really is no 'illegal' UML graphical syntax"* (sic).

Notwithstanding, we believe that UML should be a more precise language. In fact, UML's "semantics" are not really formal semantics at all, but a metamodel describing how syntactically well-formed UML diagrams should be constructed; which is not the same as giving the meaning of the UML notation [17-19].

The Discovery Method for developing object-oriented systems was first proposed in 1998 by Simons [13, 14] and revised in 2002. The method uses a simple and semantically clarified notation, based on UML, but substitutes original diagrams where this is considered necessary. The method concentrates on the technical process of analysis and design [20]. It deploys many existing analysis and design techniques selectively, restricting them to their original context and purpose, and emphasises the benefits of formal model transformations [13]. The method is consistent with the process model of OPEN [21], and has been tested in a number of industrial projects by MSc students in the University of Sheffield.

## 3. Alloy analyzer

Formal methods provide a syntactic domain (i.e. the notation or set of symbols of the method), a semantic domain (like its universe of objects), and a set of precise rules defining how an object can satisfy a specification [22]. Most formal methods are supported by one or more tools. These, based on the characteristics of each formal method, can be categorized as theorem provers and model checkers.

There is a slightly different category of tools such as the Alloy analyzer which is sometimes defined as a model finder. This kind of tool works by finding models that form counterexamples to assertions made by the user: *"Its engine takes a formula and attempts to find a model of it"* [11]. Alloy by Jackson and Paradox by Claessen et al. [23] are examples of a program that implements techniques for finding finite models based on first order logic, whilst model checking is based on temporal logic.

In Jackson's words [24] *"Alloy is an attempt to combine the best features of Z and the Object Constraint Language of UML in a lightweight notation. It takes UML's emphasis on binary relations, and the expression of constraints with sets of objects formed by 'navigations', but with Z's much simpler semantics."*

The essential idea about Alloy can be summarized as follows. We can build a micro-model with Alloy using signatures and formula paragraphs (i.e. predicates, functions, or assertions). Once the model is compiled, we can check every assertion with the intention of finding a counterexample. In other words, the Alloy analyzer looks for some instance of our micro-model that could be generated in violation of the assertions. It is for this reason that Jackson says in [25] that Alloy follows a *refutation* approach. If a counterexample is found, this means that the model is invalid. If no counterexample is found, the model may be valid, but this is not guaranteed. Alloy searches exhaustively, creating the complete state space, but within a limited scope, bounded by fixed numbers of element instances in the model [26]. If Alloy cannot find a counterexample in the scope specified, one may still exist in a larger scope. The effectiveness of this method is based on the *small scope hypothesis* [12] that states that a high proportion of specification errors tend to be found in a small scope. Alloy can also be used to search for positive instances of a model, a feature used in the work reported here.

## 4. Methodology

The abstract syntax was determined by examining each design model used in the Discovery Method in turn, then describing each model element and the constraints upon that element. Initially, there was some freedom to develop either a single abstract syntax, or a collection of syntaxes, one for each type of model.

Alloy contains certain built-in predicates that were useful when checking properties of the abstract syntax. For example, some models had the property of being directed acyclic graphs (DAGs). Provided that a relation could be constructed to generate the transitive graph, the built-in *dag()* constraint could be applied to this expression.

Successive versions of the abstract syntax specification were tested by proposing *check* assertions in Alloy, counterexamples which encoded violations of desired properties of the abstract syntax, for example that an *Object* is a composition (exclusive aggregation) of itself, recursively. When these were *checked*, Alloy would sometimes find counterexamples, indicating that the abstract syntax did not yet encode sufficient invariant properties to rule out malformed diagrams.

Later, when checking diagram instances against the abstract syntax, we switched from the refutation approach to a predicate satisfaction approach, whereby diagram instances were encoded as predicates and the Alloy analyzer had to satisfy one instance of each predicate, to indicate that a diagram was valid.

## 5. Abstract syntax

Our research is geared toward the use of Discovery (a semi-formal lightweight object-oriented method) and Alloy, used as a supporting formal method, with the aim of defining the formal representation for Discovery. In the longer term, we aim to develop a tool to support these definitions to check the consistency and completeness of our Discovery models. Figure 1 depicts the projected architecture of our model.

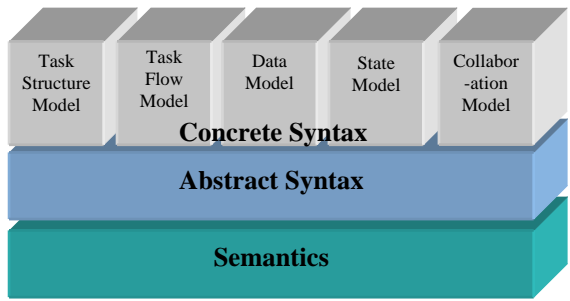

**Figure 1. Abstract syntax for the Discovery Method**

The abstract syntax for the notations of the Discovery Method has been coded in Alloy with the aim of facilitating the mapping between the notation and the semantic domain [10]. Our abstract syntax model also includes well-formedness rules or static semantics[27], which govern the correctness of Discovery models. Checking for well-formedness is traditionally made at a diagrammatical level using a BNF specification, but we work in Alloy, trying to define the whole abstract syntax and looking for an appropriate representation of the model instances of Discovery and experimenting with the model checking supported by Alloy.

At present, we have a unified abstract syntax for the five principal Discovery models (Task Structure, Task Flow, Object, State and Collaboration models), which includes well-formedness rules derived naturally from the diagram notations. We combined the abstract syntaxes for each model to have single definitions of the common elements with shared properties. A similar strategy for UML has been recommended by the 2U group in their UML 2.0 proposal [28]. Evans et al. actually propose two abstract syntaxes to support all the concrete syntax of UML [29], separating the abstract syntax describing structure from that describing behaviour. Figure 2 shows our chosen abstract syntax architecture, with four layers: the System view, the Model view, the Diagram view, and the base level for the elements of Discovery notation.

The System view gives a complete representation of a specification, formed by a collection of models in the Discovery Method. This view includes at most one model of each kind and maintains the relationships between the different models. The Model view is used to define the different models supported by Discovery. At this level, each model has $n$ diagrams and the Model view maintains the consistency between these different diagrams.

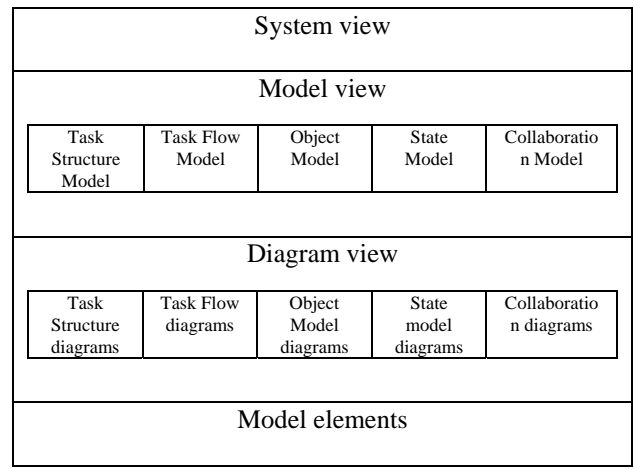

**Figure 2. General structure of the abstract syntax**

The Diagram view specifies single diagrams without concern for their interrelation, since the purpose at this level is to ensure that diagrams use the appropriate elements of Discovery's notation. The lowest level is used to specify all the relevant elements of the Discovery notation and their basic relationships.

With this layering of models and diagrams, it is possible to check, at different levels of detail:
- Each diagram separately
- Each model independently
- The whole system specification

## 6. Checking visual models

The abstract syntax model supports the definition of generic syntax constraints, together with the specific constraints relating to a particular diagram, model or system. While we may check the Alloy representation for all three views shown in Figure 2, we must always include the Diagram view, since this declares the relevant primitive elements. The strategy followed is to encode the general constraints for each type of diagram in one Alloy signature, and then to encode a specific diagram as a subtype signature in Alloy. The reasons for this are discussed below in section 7.

```
sig TaskStDiagramView
   extends DiagramView{
 task: set Task,
 goal: set Goal,
 gen: set Generalisation,
 real: set Realisation,
 agg: set Aggregation – Composition,
 comp: set Composition,
 actor: set Actor,
 obj: set Object - AssociationClass,
 parti: set Participation
}
```

**Figure 3. Task Structure diagram elements**

Figure 3 shows the signature *TaskStDiagramView*, defining the general properties of a Task Structure diagram in Alloy. This basically declares the sets of elements that can possibly be part of the diagram. The relationships among these elements are defined at the lowest level of the abstract syntax graph (see the metamodel in Appendix A).

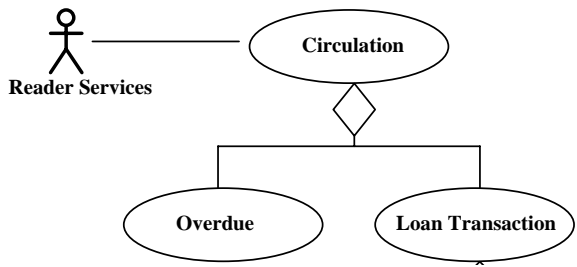

**Figure 4. Circulation Task Structure diagram**

Given the above, we may encode a specific Task Structure diagram, such as the sketch of a Library's circulation system in Figure 4. The corresponding Alloy signature s*CirculationTS*, which represents the diagram instance, is given in Figure 5. This signature extends the basic *TaskStDiagramView* signature. In the upper declaration area, the particular elements of the diagram are declared. These are all expressed in terms of diagram element types inherited from the generic signature. In the lower predicate area, constraints are defined on the declared elements. One constraint is used to specify the aggregation relationship, linking tasks to their corresponding source or target tasks in the structure. A similar constraint is created to define the participation linking the actor and the top level task.

```
sig sCirculationTS extends
   TaskStDiagramView {
 part circulationTask, overdueTask,
   loanTransactionTask: task,
 readerServicesActor: actor,
 part p: parti,
 circAgg: agg
}{
 // Aggregation
 circulationTask in circAgg.head and
   overdueTask + loanTransactionTask
   in circAgg.tail
 #circAgg.tail=2
 // participation
 circulationTask in p.tact and
   readerServicesActor in p.user
}
```

**Figure 5. Encoding the Circulation Task Structure diagram**

We may specify the abstract syntax for further Task Structure diagrams. Figure 6, for example, shows a diagram that represents an extension of the Task Structure diagram given in Figure 4. Eventually, the two independently-created diagrams should be made consistent within the same Task Structure model.

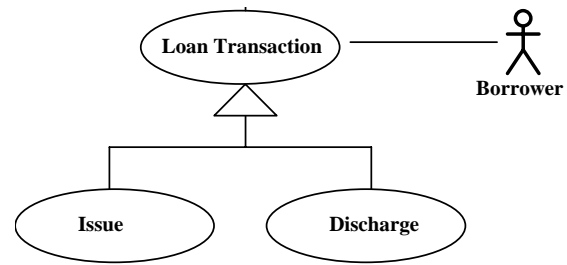

**Figure 6. Loan Transaction Task Structure diagram**

The corresponding signature *sLoanTransactionTS* is shown in Figure 7. The specification is constructed in a similar way as before, but this time describes a generalisation instead of an aggregation relationship.

With these two definitions we have enough information to check each diagram separately, to demonstrate that they each conform to the legal syntax of a Task Structure diagram. However, it is more interesting to treat them as part of the same Task Structure model and check them together. To achieve this, we must construct a new Alloy specification, representing the Model view, within which we merge the two diagrams on their common element (the *Loan Transaction* task).

```
sig sLoanTransactionTS extends
   TaskStDiagramView {
 part loanTransactionTask, issueTask,
   dischargeTask: task,
 borrowerActor: actor,
 p: parti,
 loanGener: gen
}{
// generalisation
loanTransactionTask in loanGener.head
  and issueTask + dischargeTask in
  loanGener.tail
#loanGener.tail=2
// participation
loanTransactionTask in p.tact and
  borrowerActor in p.user
}
```

**Figure 7. Encoding the Loan Transaction Task Structure diagram**

Figure 8 shows the signature *sCirculationModel*, representing a particular Task Structure model for the whole circulation subsystem, which merges the above diagrams consistently. The signature extends a generic *TaskStModel* signature (whose detail is not given here) and specifies that the diagrams *sCirculationTS* and *sLoanTransactionTS* are part of the model. All the information pertaining to the Model view is inherited from *TaskStModel* and the individual diagrams were specified above in the Diagram view, so apart from linking the diagrams to the model, we only need to assert which elements are common to both diagrams.

```
sig sCirculationModel
  extends TaskStModel {
}{
sCirculationTS in tm
sLoanTransactionTS in tm

sCirculationTS.loanTransactionTask
=
LoanTransactionTS.loanTransactionTask

sCirculationTS.task
&  sLoanTransactionTS.task
=
sLoanTransactionTS.loanTransactionTask
}
```

**Figure 8. Encoding the Task Structure model**

Having defined a particular Task Structure model consisting of two Task Structure diagrams, it is possible to check the consistency of these against the rules of the abstract syntax. If Alloy cannot find a valid instance, this will mean that our model does not conform to all the syntax constraints defined for the Discovery notation. Figure 9 illustrates the Alloy code that is executed to validate our model. This consists of a dummy predicate *circulationModel()* which is run for an exactly-specified scope, within which Alloy must find all the elements of the model. The scope is an enumeration of each element and relationship used in the model under test. In a smaller scope Alloy cannot generate a valid instance, whilst in a larger scope Alloy will create additional elements, making the instance valid with the whole abstract syntax, but not equivalent to our model. Indicating the exact scope is necessary if satisfaction is to be interpreted as validating the model. But this also has the useful effect of limiting the state space searched by Alloy for a valid instance.

```
pred circulationModel(){}
run circulationModel
 for 1 but
 exactly 1 Model,
  exactly 1 TaskStModel,
    exactly 1 sCirculationModel,
 exactly 2 DiagramView,
    exactly 2 TaskStDiagramView,
      exactly 1 sLoanTransactionTS,
      exactly 1 sCirculationTS,
 exactly 4 Relationship,
    exactly 2 Structure,
      exactly 1 Generalisation,
      exactly 1 Aggregation,
    exactly 2 Participation,
 exactly 7  Node,
    exactly 5 StateAndTask,
      exactly 5 TaskActivity,
        exactly 5 Task,
    exactly 2 Actor,
 0 Transition,
 0 TaskFlowElement,
 0 Member
```

**Figure 9. Empty predicate and exact scope specified for the run command**

When the above *run* command is executed, Alloy finds the unique instance, indicating that our example is in fact consistent with the Discovery abstract syntax. What Alloy does is to satisfy the empty predicate (a trivial task in itself) in conjunction with making the particular diagram and model specifications consistent with the general syntax specifications, within a scope that only has one possible solution, if any. Alloy presents its result either as a graph of linked signature instances (similar to the metamodel graph in Appendix A), or as a browseable tree (as shown in Figure 10).

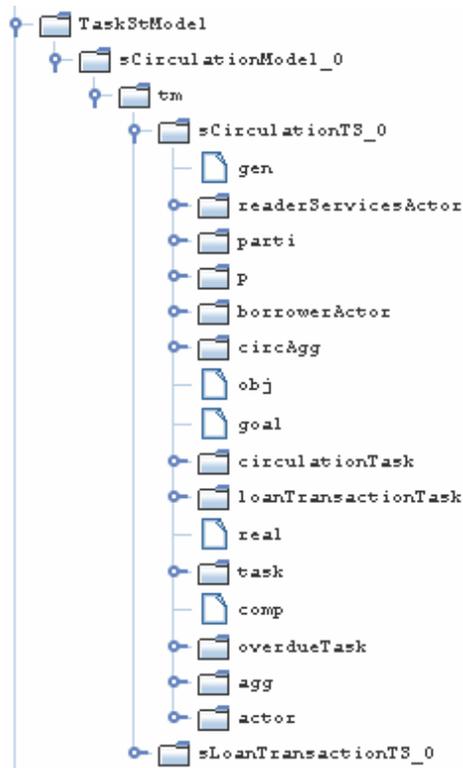

**Figure 10. Solution generated by Alloy**

Figure 10 shows the tree view generated by Alloy for the example presented above, whose structure we can inspect interactively if we want to examine the result. The fact that Alloy finds an instance at all demonstrates that the example is valid. If no result is returned, this means that the tested model is invalid.

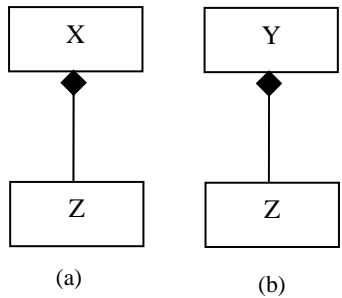

**Figure 11. Two diagrams creating an inconsistent Data Model**

Figure 11 illustrates a second interesting example, for which we would expect no consistent solution to be found by Alloy. It is possible to verify that the individual exemplar diagrams (a) and (b) are syntactically correct in the Diagram view, but when both diagrams are included within the same Data Model in the Model view, Alloy cannot find a valid instance. This is because the Z class is defined as a component of two different classes in the same model, something which violates the specification for a UML *composition*, which requires the composed elements to be uniquely-owned parts of the whole.

## 7. Evaluating Alloy

While Alloy is very effective in modelling and analysing simple, lightweight formal specifications written in a Z-like style, we found that it is more difficult to use as the basis for model checking the syntax and static semantics of a design notation. At various times, we found we were forced into work-arounds to constrain the searching behaviour of the analyzer. The following gives a flavour of some of our unexpected discoveries while modelling in Alloy.

Initially, we developed a separate abstract syntax for each type of model used in the Discovery Method. So, for example, the Task Structure Model had distinct generalisation and aggregation relationships from those in the Data Model, although in the Discovery Method these are each single kinds of relationship, with a uniform semantics across all model types. This meant that the Alloy signatures for *Generalisation* and *Aggregation* were short and the scopes, within which model instances were checked, were quite small. However, when models of different types were combined, this required a set of translations from one abstract model syntax to another.

In the second version, we unified all the abstract syntaxes for the different model types, such that a single *Aggregation* relationship existed for all types of model. This was more in keeping with the philosophy of the Discovery Method. However, the Alloy signature for *Aggregation* was made more complicated by the need to assert extra constraints that it either related two *Tasks*, or two *Objects* and not one of each. Alloy lends itself to creating hierarchies of disjoint subtypes in its abstract syntax, using the *extends* notation. This initially fostered a meta-modelling style of construction, whereby all syntax elements descended from a common *ModelElement* root, similar to the MMF [4]. However, this had the unexpected consequence of requiring vastly larger scopes within which to search for model instances, since Alloy interprets all scope instructions as relating to the base instances in any tree. As a necessity, the syntax tree was broken down into a series of shorter trees (see Appendix A), losing the abstraction over all model elements.

Once the abstract syntax had been fully validated using *check* assertions, we developed Alloy representations of diagram instances. Initially, a diagram instance was represented as an Alloy *predicate*, to be evaluated against generated instances of the abstract syntax. Eventually, this proved to be unwieldy, requiring the repetition of constraints whenever a part of the predicate referenced the same sub-elements in the diagram. In the second version, diagram instances were constructed as subtypes of the canonical abstract syntax types, a strange but economical encoding, which avoided such repetition of constraints. The eventual predicate to check was then trivial (empty), since all the analyser had to do was find one instance of the diagram itself. To control this, we set the scope to generate exactly one instance of each model element present in the diagram, a brute force approach to ensure that Alloy did not over-generate elements of the diagram. If the search to satisfy the trivial predicate generated a single matching instance of the diagram, then this represented success in satisfying the abstract syntax. We were able to find single instances of consistently-merged diagrams. The attempt to find an instance of mutually inconsistent diagrams failed, as expected, although no useful information could be reported about the detected inconsistency.

## 8. Conclusions

In this paper we have presented our experiences using the Alloy analyzer to check an abstract syntax for the notation of the Discovery Method. We described how we used different approaches to design the abstract syntax and to represent the diagram instances in Alloy, commenting on the naturalness, or otherwise, of the chosen encodings.

We illustrated a complete example of a valid model for Discovery (a Task Structure Model) and the result generated by Alloy, showing that the basic approach is feasible. The time taken to validate larger models with an exact scope is in the order of minutes. We also illustrated a counter-example of an invalid model (a Data Model), for which Alloy correctly found no instance.

Additionally, we gave our impressions of Alloy as a candidate tool for checking the consistency of multiple diagrams in software engineering notations. We feel that this is perhaps not an ideal deployment of Alloy. The searching behaviour of the constraint solver had to be carefully controlled. We were forced to abandon the notion of a single hierarchy of model elements in the abstract syntax specification, since this gave rise to underconstrained instance generation.

## Appendix A. Abstract syntax metamodel

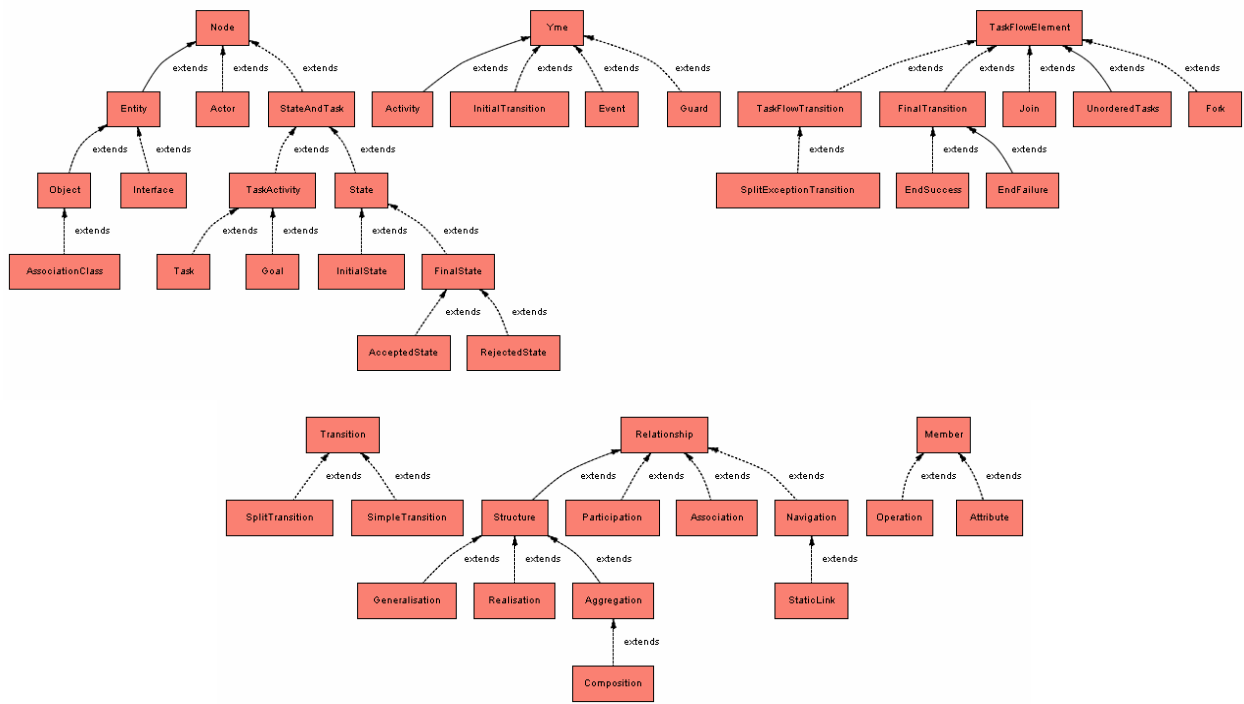